\documentclass[twocolumn]{aa}
\usepackage{rotating}
\usepackage{graphicx}
\usepackage{sidecap}
\usepackage{subcaption}
\usepackage{url}
\usepackage[colorlinks=true]{hyperref}
\hypersetup{citecolor= blue, linkcolor=blue, urlcolor=blue}

\usepackage{microtype} % improve text space

\usepackage{txfonts}
\usepackage{float}
\usepackage{natbib}
\bibpunct{(}{)}{;}{a}{}{,}
\DeclareUnicodeCharacter{00A0}{~}
\usepackage[us,12hr]{datetime}
\usepackage{ulem}

\newcommand{\kms}{\,km\,s$^{-1}$}
\newcommand{\los}{\scriptscriptstyle \rm{LOS}}

\newcommand{\cair} {\ion{Ca}{ii}~8542\,\AA\:}
\newcommand{\cak} {\ion{Ca}{ii}~K}
\newcommand{\Halpha}{H$\alpha$}
\newcommand{\CaII} {\ion{Ca}{ii}}
\newcommand{\MgII} {\ion{Mg}{ii}}

\usepackage{cancel}
\usepackage{color}

\definecolor{deepmagenta}{rgb}{0.8, 0.0, 0.8}

\newcommand{\mparallel}{{\mkern2.5mu\vphantom{\perp}\vrule depth 0.5pt\mkern2.5mu\vrule depth 0.5pt\mkern3mu}}

\patchcmd{\thebibliography}{\clubpenalty4000}{\clubpenalty10000}{}{}
\patchcmd{\thebibliography}{\widowpenalty4000}{\clubpenalty10000}{}{}

\begin{document}

    % \title{Chromospheric heating driven by cancellation of flux emergence\\
    % \subtitle{Multi-line inversion of spectropolarimetric SST data}
    
    \title{An observationally-constrained model of strong magnetic reconnection in the solar chromosphere}

    \subtitle{Atmospheric stratification and estimates of heating rates}

   \author{C. J. D\'iaz Baso
          \inst{1}
          \and
          J. de la Cruz Rodr\'iguez
          \inst{1}
          \and
          J. Leenaarts
          \inst{1}
          }

   \institute{Institute for Solar Physics, Dept. of Astronomy, Stockholm University, AlbaNova University Centre, SE-10691 Stockholm, Sweden
              \email{carlos.diaz@astro.su.se}
             }

%   \date{Received Month Day, 2019; accepted Month Day, 2019}
   \date{Draft: compiled on \today\ at \currenttime~UT}

   \authorrunning{D\'iaz Baso et al.}
    % \titlerunning{}
% \abstract{}{}{}{}{}
% 5 {} token are mandatory

\abstract
{The evolution of the photospheric magnetic field plays a key role in the energy transport into the chromosphere and the corona. In active regions,  newly emerging magnetic flux interacts with the pre-existent magnetic field, which can lead to reconnection events that convert magnetic energy into thermal energy.}
{We aim to study the heating caused by a strong reconnection event that was triggered by magnetic flux cancellation.}
{We use imaging-spectropolarimetric data in the \ion{Fe}{i}~6301\& 6302\,\AA, \ion{Ca}{ii}~8542\,\AA\ and \ion{Ca}{ii}~K   {spectral lines} obtained with the CRISP and CHROMIS instruments at the Swedish 1-m Solar Telescope. This data was inverted with the STiC code by performing multi-atom, multi-line non-LTE inversions. These inversions yielded a three-dimensional model of the reconnection event and surrounding atmosphere, including temperature, velocity, microturbulence,   {magnetic field}, and the radiative loss rate.}
{The model atmosphere shows the emergence of magnetic loops with a size of several arcsecs into a pre-existing predominantly unipolar field. Where the reconnection region is expected to be, we see an increase in the chromospheric temperature of roughly 2000~K as well as bidirectional flows of the order of 10~\kms\ emanating from there. We see bright blobs of roughly 0.2~arcsec in diameter in the \ion{Ca}{ii} \,K moving at a plane-of-the-sky velocity of the order of 100~\kms\ and a blueshift of 100~\kms, which we interpret as ejected plasmoids from the same region. This scenario is consistent with theoretical reconnection models hence providing evidence of a reconnection event taking place. The chromospheric radiative losses at the reconnection site are as high as 160~kW~m$^{-2}$, providing a quantitative constraint on theoretical models that aim to simulate reconnection caused by flux emergence in the chromosphere.}
{}

  %
 % During the cancellation an enhancement in the transverse magnetic field of 1kG is also found in the chromosphere. 
  %
 % Radiative losses computed from the inferred model reach values around 100-160\,kW\,m$^{-2}$ indicating the large energy supplied during this event to sustain the chromosphere at such temperatures for more than two hours ($\sim 5\cdot10^{27}$\,erg). 
  %
 % This detailed characterization of our observations provides important constraints on theoretical models of chromospheric heating mechanisms.
 % }
 
%   \abstract
  % context heading (optional)
  % {} leave it empty if necessary
%   {}
  % aims heading (mandatory)
%   {}
  % methods heading (mandatory)
%   {}
  % results heading (mandatory)
%   {}
  % conclusions heading (optional), leave it empty if necessary
%   {}


   \maketitle

% A maximum of 6 keywords should be listed after the abstract.
% https://www.aanda.org/index2.php?option=com_content&task=view&id=170&Itemid=256

%%%%%%%%%%%%%%%%%%%%%%%%%%%%%%%%%%%%%%%%%%%%%%%%%%%%%%%%%%%%%%%%%%%%%%%%%
\section{Introduction}\label{sec:intro}
%%%%%%%%%%%%%%%%%%%%%%%%%%%%%%%%%%%%%%%%%%%%%%%%%%%%%%%%%%%%%%%%%%%%%%%%%

The evolution of the photospheric magnetic field plays a key role in the energy transport into the chromosphere and the corona. In active regions, the emergence of new magnetic flux leads to interactions with the pre-existing magnetic field, releasing magnetic energy, heating the solar atmosphere, and accelerating the solar plasma. Signatures of such activity are commonly present in observations: from compact brightenings such as Ellerman bombs (EBs), Ultraviolet bursts (UVBs), and microflares (MFs) to big eruptive flares or X-ray jets \citep{Archontis2012, Eric2014, Young2018}. These phenomena are thought to be manifestations of magnetic reconnection that occurred at different plasma conditions. Waves also have an important role in the transfer of energy as those jets develop into shock waves as they propagate into the corona \citep{Yokoyama1995}.

A particular set of these small-scale reconnection events are induced by the interaction of opposite-polarity magnetic elements, which can leave observational imprints in the photosphere and the chromosphere. Among these phenomena, EBs \citep{Ellerman1917} are observed as rapid flames visible in the wings of H$\alpha$ or \ion{Ca}{ii}~8542\,\AA\ tracing magnetic reconnection in the lower photosphere \citep{Vissers2015}. UVBs \citep{Peter2014Sci} are compact transient brightenings observed in UV bands, with very complex \ion{Si}{iv} profiles, and do not generally show a co-spatial brightening in the coronal channels of the Solar Dynamics Observatory (SDO) Atmospheric Imaging Assembly (AIA) instrument. Similar targets have been studied with recent simulations \citep{Danilovic2017, Hansteen2017, Hansteen2019,2020arXiv201015946L} and observations \citep{Socas2006, Toriumi2017, 2017A&A...598A..33L, Salvo2018A, Smitha2018, Vissers2019, Yadav2019, Ortiz2020, Silva2020}. Microflares or stronger events show a visible response in coronal channels and they would originate from or extend into higher layers. Bi-directional flows at these locations are commonly reported in both observations and simulations \citep{Peter2014Sci, Vissers2019, Hansteen2017}. They usually have time scales of a few minutes \citep{Vissers2019, Smitha2018}, however, they can live up to several hours \citep{Salvo2018A, Chitta2017}. Therefore, differences in their lifetime, size, location, etc., are used to distinguish them. This classification is even more difficult when several of these phenomena can occur co-temporarily due to simultaneous reconnection at different heights \citep{Ortiz2020, Hansteen2019}. They sometimes are associated with a cool counterpart known as surges, which are observed as darkenings in images taken in the \Halpha\ blue/red wings with large velocities \citep{Yokoyama1995, Nobrega2016}.

Given their transient small-scale nature, high-spatiotemporal-resolution spectropolarimetric observations are also needed to discern the mechanisms involved in the reconnection. Observations and simulations of similar explosive events \citep[e.g.;][]{Innes2015, 2015ApJ...799...79N, Rouppe2017, Nobrega2016, Peter2019} have found very small-scale plasmoids that are ejected at high speeds after the fragmentation of the current sheet (tearing instability) during magnetic reconnection that enable fast energy release. In contrast, other studies propose the Kelvin-Helmholtz instability to generate a turbulent velocity field which finally triggers fast reconnection \citep[e.g.;][]{Lazarian99ApJ, Kowal2009, Jeffrey2018, Chitta2020}.

Chromospheric energy deposition can be estimated through the radiative losses, which account for the energy that must be replenished at any given time by heating mechanisms. \citet{Leenaarts2018} investigated the chromospheric heating in a strong flux emerging region. They used the \cak\ brightness as a proxy of the radiative losses in the chromosphere finding a good correlation with the total linear polarization, which originated from large-scale loop-like structures. However, they did not attempt estimating radiative losses from their observations. To our knowledge, spatially-resolved estimates of chromospheric radiative losses have not been   {carried out} in emerging flux regions.

In this study, we present an analysis of chromospheric heating induced by magnetic flux cancellation rooted in the photosphere. Our target is larger in size than typical EBs or UVBs and more similar to that observed by \citet{Smitha2018} or \citet{Salvo2018A}. We have reconstructed 3D models from NLTE inversion of high-resolution spectropolarimetric data acquired in the \ion{Ca}{ii}~K, \ion{Ca}{ii}~8542\,\AA\ and \ion{Fe}{i}~6301\& 6302\,\AA\ lines. Our study focuses on the overall atmospheric stratification of temperature, line-of-sight velocity, and magnetic field vector. In addition, we estimate for the first time the total radiative losses of such a strong magnetic reconnection event, allowing us to calculate the energy needed to sustain the high inferred temperatures and compare it with the magnetic energy available. We also present an analysis of the evolution of the polarization signals, as well as the detection of plasmoids during the time series.

% \clearpage
%%%%%%%%%%%%%%%%%%%%%%%%%%%%%%%%%%%%%%%%%%%%%%%%%%%%%%%%%%%%%%%%%%%%%%%%%
\section{Observations}\label{sec:observation}
%%%%%%%%%%%%%%%%%%%%%%%%%%%%%%%%%%%%%%%%%%%%%%%%%%%%%%%%%%%%%%%%%%%%%%%%%

%%%%%%%%%%%%%%%%%%%%%%%%%%%%%%%%%%%%%%%%%%%%%%%%%%%%%%%%%%%%%%%%%%%%%%%%%
\subsection{Target and observational setup}\label{sec:setup}
%%%%%%%%%%%%%%%%%%%%%%%%%%%%%%%%%%%%%%%%%%%%%%%%%%%%%%%%%%%%%%%%%%%%%%%%%

Our target develops in the active region NOAA 12593, which started emerging on 2016-09-18 around 04:00\,UT.  We observed it on 2016-09-21 recording three time-series of around 11, 17, and 21\,min at 11:00h, 12:15h, and 12:38h\,UT, respectively. The observations were performed using the CRisp Imaging SpectroPolarimeter \citep[CRISP;][]{Scharmer2008} and the CHROMospheric Imaging Spectrometer \citep[CHROMIS;][]{Scharmer2017} instruments at the Swedish 1-m Solar Telescope \citep[SST;][]{Scharmer2003}. The active region was located close to solar disk center, with solar coordinates at the beginning of the observations  (X, Y) = (400\arcsec, 25\arcsec) and an observing angle $\mu$ = 0.91.

The \ion{Fe}{i} 6301.5\,\AA\ and the 6032.5\,\AA\ data were recorded with the CRISP instrument at 16 wavelength positions. The \ion{Ca}{ii} 8542\,\AA\ data were also recorded with CRISP at 21 line positions:  two wavelength positions at $\pm$1.7\,\AA\ from the line core, and 19 positions spaced evenly between $\pm$0.765\,\AA. Both datasets were recorded in full-Stokes mode. The H$\alpha$ line was sampled in  16  wavelength positions non-equidistantly spaced between $-$2.2\,\AA\, and $+$1.3\,\AA\ from the line core. Datasets in the \ion{He}{i}~5875\,\AA\ (D$_3$) line were also acquired, but they are not used in this study.  A complete scan of the \ion{Fe}{i}, \ion{Ca}{ii} and \ion{H}{i} lines required 37\,s. In the first and third scan \ion{He}{i} D$_3$ was acquired instead of the \ion{Fe}{i} lines, with a cadence of 43\,s. The \ion{Ca}{ii}~K dataset was recorded using the CHROMIS instrument. The latter was sampled using two-wavelength positions at $\pm$1.33\,\AA\ from the line core, 19 positions spaced evenly between $\pm$0.7\,\AA\ around the line center (i.e. 3933.682\,\AA), as well as a single wavelength point in the pseudo-continuum at 4001\,\AA. The cadence of the CHROMIS data is approximately 9\,s. Unfortunately, this event has not been observed by RHESSI or IRIS.

\begin{figure}
\centering
\includegraphics[width=\linewidth]{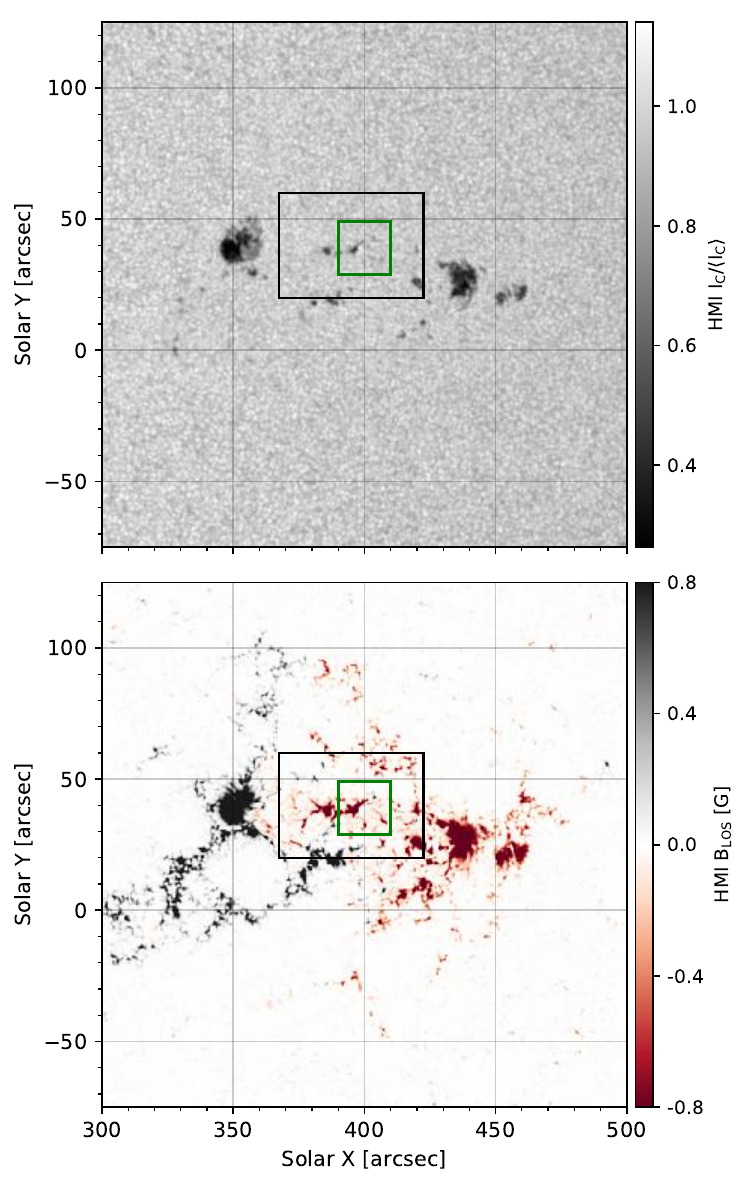} %width=\linewidth
\caption{SDO/HMI continuum image and SDO/HMI line-of-sight magnetogram of the active region NOAA 12593 at 12:15h\,UT. The black square   {contour} outlines the FOV covered by the SST and the smaller green square the region-of-interest (ROI).}
\label{fig:sdomaps}
% /scratch/carlos/CRISP/CONTEXT/SDOAIA$ python Completa_v3.py
\end{figure}

\begin{figure*}
\centering
\includegraphics[width=\linewidth]{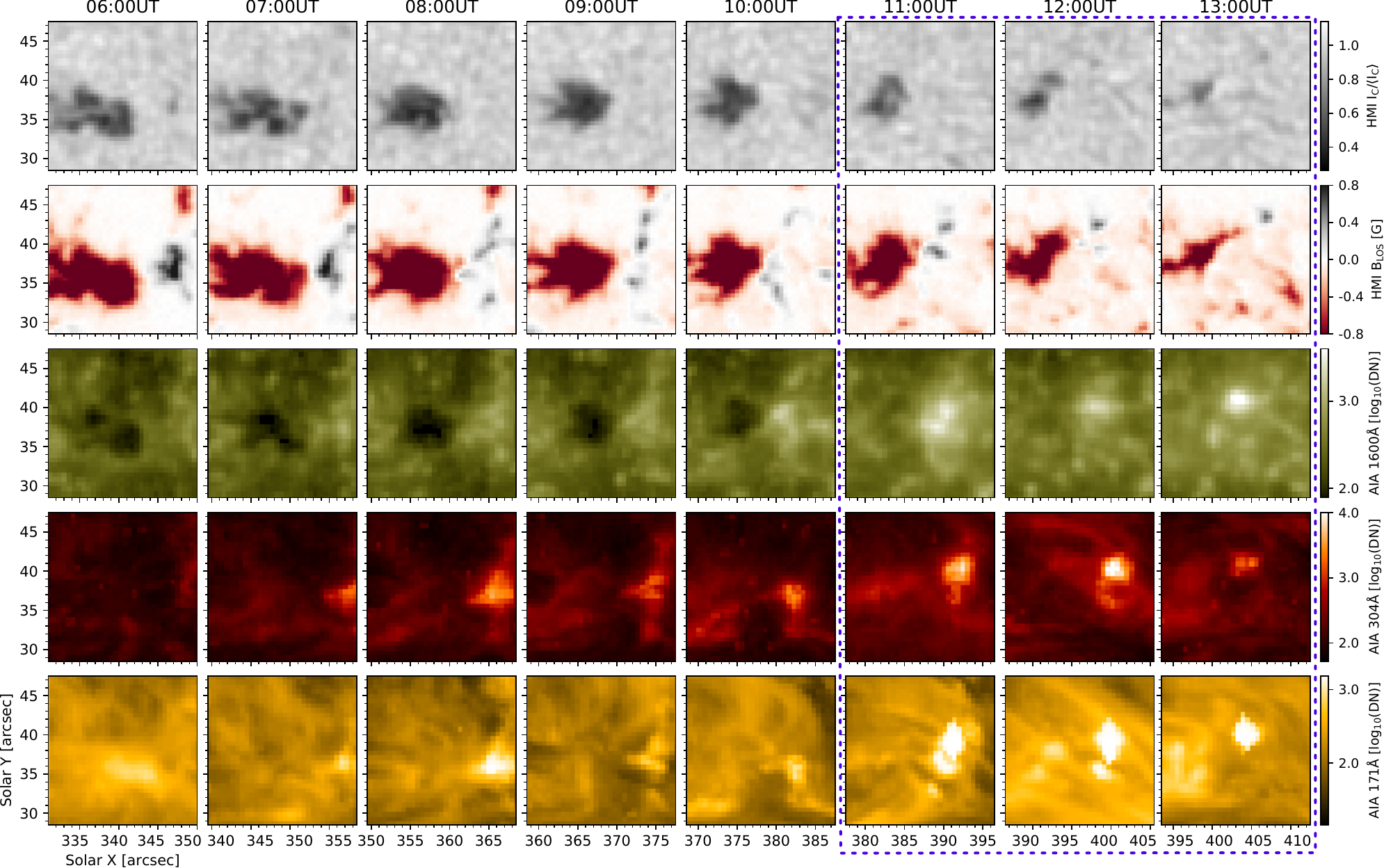}
\caption{Time evolution of the ROI (green region shown in Fig.~\ref{fig:sdomaps}) as seen from SDO-HMI and SDO-AIA channels. The purple dashed-line square indicates the temporal window where we have SST observations.}
\label{fig:evolution}
% /scratch/carlos/CRISP/CONTEXT/SDOAIA$ python miniEVOLUTION_v3.py
\end{figure*}

%%%%%%%%%%%%%%%%%%%%%%%%%%%%%%%%%%%%%%%%%%%%%%%%%%%%%%%%%%%%%%%%%%%%%%%%%
\subsection{Data reduction}\label{sec:reduction}
%%%%%%%%%%%%%%%%%%%%%%%%%%%%%%%%%%%%%%%%%%%%%%%%%%%%%%%%%%%%%%%%%%%%%%%%%

Both data sets were reduced using the CRISPRED and CHROMISRED pipelines as described by \citet{delaCruz2015} and \citet{Lofdahl2018}, using the Multi-Object-Multi-Frame-Blind-Deconvolution image restoration technique \citep[MOMFBD;][]{vanNoort2005} to remove the atmospheric distortion. Residual rubber-sheet seeing distortions were compensated as proposed by \cite{Henriques2012A}.

The CRISP and CHROMIS datasets were accurately aligned by scaling up the CRISP pixel scale to that of CHROMIS (from $0.059$\arcsec\ to $0.038$\arcsec). Given that the temporal cadence of the CHROMIS data is higher, CRISP images were interpolated to the CHROMIS cadence using nearest-neighbor interpolation in the temporal dimension. As a final step, we performed an intensity calibration of the data by scaling the intensity in a quiet-sun area in the observations to the values of a reference atlas taking into account the observed heliocentric angle following \cite{Neckel1984} and \cite{Allen}\footnote{These routines are available in the Python repository\\ \url{https://github.com/ISP-SST/ISPy}}.

After the standard data reduction, a final inspection showed some residual artifacts that were removed in the following way: polarized interference fringes of short period in the monochromatic images were isolated in the Fourier space and filtered out. Long-period fringes identified at nearly-continuum wavelengths in the polarization signals were subtracted to the rest of the wavelength points by removing their contribution (scalar projection). For a more detailed explanation of those procedures, see \cite{Diaz2019_8542FIL}. To reduce the noise in the polarization signals, we have used the neural network presented in \cite{DiazBaso2019_denoiser} on Stokes $Q$, $U$ and $V$. We estimate that the final data has a polarimetric accuracy of $6\cdot10^{-4} I_c$.

%%%%%%%%%%%%%%%%%%%%%%%%%%%%%%%%%%%%%%%%%%%%%%%%%%%%%%%%%%%%%%%%%%%%%%%%%
\subsection{Context data of SDO}\label{sec:context}
%%%%%%%%%%%%%%%%%%%%%%%%%%%%%%%%%%%%%%%%%%%%%%%%%%%%%%%%%%%%%%%%%%%%%%%%%

We used data provided by the Atmospheric Imaging Assembly \cite[AIA;][]{aia2012} and the Helioseismic and Magnetic Imager \cite[HMI;][]{hmi2012} to put the observations into context. HMI and AIA are both onboard NASA's Solar Dynamics Observatory \cite[SDO;][]{sdo2012}. The active region at the time of our observations (around 12:15\,UT) is displayed in Fig.~\ref{fig:sdomaps}. The field-of-view (FOV) of the SST (drawn as a black square) was surrounded by two small sunspots of different polarities (see the magnetogram at the bottom of Fig.~\ref{fig:sdomaps}). Inside the SST FOV, an emerging flux region was observed at that time near the location of some pre-existent pores, indicated with a green square. This is our region of interest (ROI).
    
We show the history of our ROI as seen in SDO observations in Fig.~\ref{fig:evolution}. The simultaneous continuum intensity and magnetograms show the evolution of the corresponding phenomena in the photosphere. Before 06:00\,UT there was only a pore of negative polarity at the center of the ROI. A new magnetic concentration of positive polarity of around 5\arcsec\ emerges at the surface around 06:00\,UT (visible in the magnetogram of the same figure). After that, they converged   {towards} each other. They became smaller until the new magnetic concentration almost disappears after 13:00\,UT (rightmost column). The total magnetic flux of both features decreases over time by a roughly similar amount. We estimated the total canceled flux around $4\cdot10^{19}$\,Mx, and the cancellation rate around 10$^6$\,Mx\,s$^{-1}$. The AIA panels (lower rows of Fig.~\ref{fig:evolution}) show the evolution of this event in the chromosphere and corona. After the magnetic concentrations are in contact, some brightenings are visible at the location of the flux cancellation (around 08:00UT) that became brighter by one to two orders of magnitude at the end of the process around 11:00-13:00UT (coinciding in time with our observations highlighted with a purple dashed square).

Some form of magnetic reconnection is expected in this situation of a new flux   {emerging} region inside a pre-existing active region. The energy released during this interaction may be responsible for the observed chromospheric and coronal brightening.

\begin{figure*}
\centering
\includegraphics[width=0.96\linewidth]{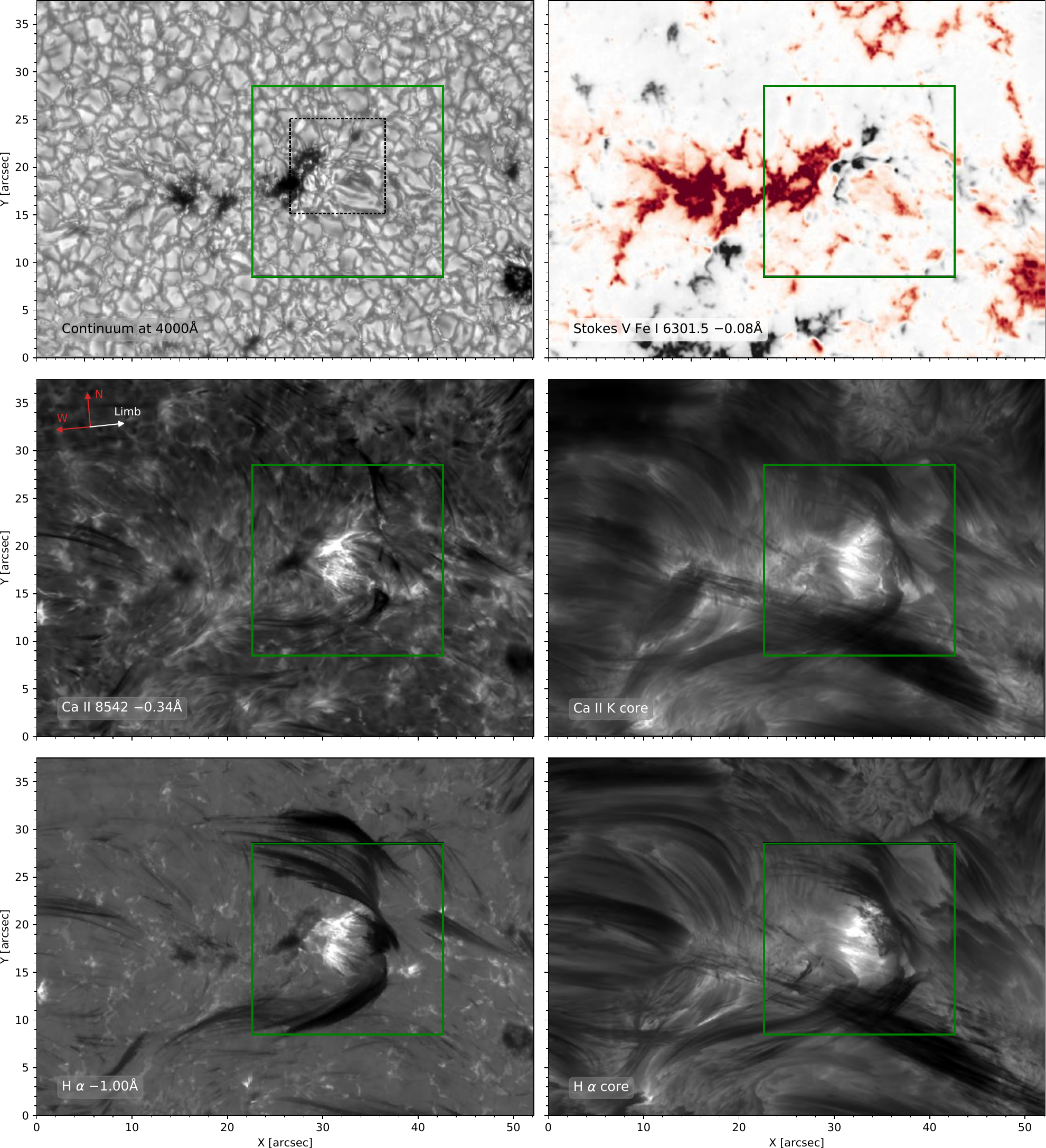}
\caption{Overview images which overlap between the two instruments at four different wavelengths taken at 12:15UT. The red arrows point to Solar North and West, while the white arrows indicate the direction to the closest limb. The green square indicates our ROI which has been analyzed in detail with the inversion code. The black dashed square is a close-up analyzed in Sect.~\ref{sec:plasmoid}.}
\label{fig:mosaico}
% /scratch/carlos/CRISP/1COMPOSITE$ python mosaico_grande.py
% /scratch/carlos/CRISP/CONTEXT/LOCATION/ python main.py
% scratch/carlos/CRISP/1COMPOSITE$ python mosaico_grande_v4.py
\end{figure*}

%The imprint is mostly visible in all the channels with an intensity of about two orders of magnitude higher than the typical values, indicating the high temperatures reached in this region. 

%%%%%%%%%%%%%%%%%%%%%%%%%%%%%%%%%%%%%%%%%%%%%%%%%%%%%%%%%%%%%%%%%%%%%%%%%
\subsection{SST observations}\label{sec:sstdata}
%%%%%%%%%%%%%%%%%%%%%%%%%%%%%%%%%%%%%%%%%%%%%%%%%%%%%%%%%%%%%%%%%%%%%%%%%

Although the brightenings are visible in the HMI and AIA channels, due to their moderate resolution ($\sim$$1.2$\arcsec\ in HMI and $\sim$$1.6$\arcsec\ in AIA) it is not possible to study the internal structure of the region. In the much higher spatial resolution SST observations (close to 0.1\arcsec at 4000\,\AA), the region shows a very complex morphology. The SST FOV is indicated in Fig.~\ref{fig:mosaico}, where we highlight in green the same ROI defined in previous figures. The photospheric images at SST resolution contain elongated granules generated during the flux emergence as shown in the upper-left panel of Fig.~\ref{fig:mosaico}. We also identify the complex shape of the different magnetic polarities during the interaction, as shown in the Stokes $V$ at the wing of the \ion{Fe}{i} line in the upper-right panel of the same figure.

In the chromosphere, the brightenings start to be visible in the wings and core of \ion{Ca}{ii}~8542\,\AA, \cak\ and \Halpha\ as bright fibrils directed to the East, emerging apparently from the interface between both polarities with widths of 80-200~km (bottom panels of Fig.~\ref{fig:mosaico}). These values are estimated using the \ion{Ca}{ii}~K images by measuring the full-width at half-maximum (FWHM) of the intensity profile across the fibrils. Above those brightenings, we observe dark jets that are apparently escaping from the tail of the brightenings but they bend backward, moving towards the opposite direction (West). They are better visible in the blue wing of \Halpha\ with a projected length of approximately 10~Mm. Their detection in the blue wing and not in the red wing implies strong upflows of at least 50\,\kms\ (points further in the wings also shows similar features). These dark jets could then be identified as surges, coexisting with the hot fibrils \citep{Nobrega2017}.

The intense chromospheric emission is found during the entire time-series, of almost 2 hours. Although the evolution in the photosphere is very slow and the presence of chromospheric emission is relatively stable in time, the \ion{Ca}{ii} images also show a dynamic response with rapidly moving brightenings that evolve on timescales faster than the CHROMIS cadence of 9 seconds.

% These chromospheric and coronal imprints suggest an extended enhancement of temperature from the upper photosphere/lower chromosphere to higher up in the corona.

% \begin{itemize}\setlength\itemsep{1em}
% \end{itemize}

%%%%%%%%%%%%%%%%%%%%%%%%%%%%%%%%%%%%%%%%%%%%%%%%%%%%%%%%%%%%%%%%%%%%%%%%%
\section{Data analysis}\label{sec:analysis}
%%%%%%%%%%%%%%%%%%%%%%%%%%%%%%%%%%%%%%%%%%%%%%%%%%%%%%%%%%%%%%%%%%%%%%%%%

%%%%%%%%%%%%%%%%%%%%%%%%%%%%%%%%%%%%%%%%%%%%%%%%%%%%%%%%%%%%%%%%%%%%%%%%%
\subsection{Non-LTE inversions}\label{sec:inversion}
%%%%%%%%%%%%%%%%%%%%%%%%%%%%%%%%%%%%%%%%%%%%%%%%%%%%%%%%%%%%%%%%%%%%%%%%%

% 3D structure of the
We estimated the   {model} atmosphere for the ROI through non-Local Thermodynamic Equilibrium (non-LTE or NLTE) inversions. We performed an inversion of the Stokes profiles in the \ion{Fe}{i}~6301.5 \& 6302.5 \AA, \ion{Ca}{ii}~8542~\AA\ and \ion{Ca}{ii}~K lines, simultaneously, using the parallel non-LTE STockholm Inversion Code\footnote{\url{https://github.com/jaimedelacruz/stic}} \cite[STiC;][]{delaCruz2016,delaCruz2019_STiC}. This code   {is built on top of} an optimized version of the RH code \citep{Uitenbroek2001}. It makes use of a fast angular approximation to include the effect of partial redistribution \citep[PRD;][]{Leenaarts2012} which is necessary   {for a realistic modelling of} the \ion{Ca}{ii}~K line formation. Electron densities were derived consistent with non-LTE hydrogen ionization, by iteratively solving the statistical equilibrium equations   {whilst enforcing charge conservation}. The polarized radiative transfer equation was integrated using an efficient cubic DELO-Bezier formal solver \citep{2013ApJ...764...33D}. 
  {A three-dimensional model atmosphere is obtained assuming 1D plane-parallel atmospheres along each line-of-sight. This approximation is usually referred to as 1.5D modeling}.

STiC iteratively adjusts the physical parameters of a model atmosphere, such as the temperature, line-of-sight velocity, magnetic field vector, and microturbulence to find a synthetic spectrum that reproduces the observed profile. The stratifications of density and gas pressure are computed by assuming hydrostatic equilibrium (HE). The physical parameters of the model atmosphere are given as functions of the optical depth scale at 5000\,\AA {, hereafter $\log(\tau)$}. The physical parameters are modified at specific node locations, followed by an interpolation to all other depth points. The merit function that accounts for the likelihood between the synthetic and observed spectra includes an additional regularization term that ensures a smooth behavior in the atmosphere   {also} when a physical parameter has many degrees of freedom (i.e., nodes in the stratification).

We treated the \ion{Ca}{ii} atom in non-LTE with the \cak\ line in PRD, the \cair line in complete frequency redistribution (CRD), and the \ion{Fe}{i} lines as well in non-LTE in case the complex atmosphere could affect their formation \citep{Smitha2020}. We initialize the model atmosphere from the FAL-C   {model} \citep{Fontenla1993} by interpolation to 54 depth points from $\log(\tau)=-7.0$ to $\log(\tau)=+1.0$. We have used an enhanced gas pressure at the upper boundary that accounts for the typical values of active regions for all pixels \citep[P$_{\rm top}$= 1.0 dyn cm$^{-2}$;][]{delaCruz2019_STiC}. The inversions were run in multiple cycles with spatial smoothing between cycles, with the general approach being to use fewer nodes in the firsts cycle to obtain the general trend and more nodes in the subsequent cycles to get a more detailed atmospheric structure. We have initialized the magnetic field by using a Milne-Eddington inversion\footnote{\url{https://github.com/cdiazbas/LMpyMilne}} of the \ion{Fe}{i} lines. The inversion strategy is summarized in Table~\ref{tab:cycles}.  To have an estimate of the uncertainty of our model after the inversion, we have used Eq.~42 of \cite{delToro2016} which takes into account both the discrepancy between the observed and the synthetic profiles and the effective sensitivity of the spectral lines to the underlying parameters, except that we only introduce perturbations at the node location. The uncertainty is calculated at the node location and then interpolated to the whole stratification for better visualization.

% where preventively also treated in non-LTE
% We have used $\log(\tau)$ instead of the usual $\log(\tau_{500})$ to simplify the notation.

%   {A three-dimensional model atmosphere is obtained assuming 1.5D plane-parallel atmospheres where each pixel is independent of each other.}

% slowly increase the number of nodes from cycle to cycle to first capture a general trend and gradually move towards a more detailed atmospheric structure
%%%%%%%%%%%%%%%%%%%%%%%%%%%%%%%%%%%%%%%%%%%%%%%%%%%%%%%%%%%%%%%%%%%%%%%%%
\begin{table}
\caption{Number of nodes used in each inversion cycle.}
\label{tab:cycles}
\centering
\begin{tabular}{c c c c c c c}
\hline\hline
   Cycle/Parameter & $T$ & $v_{\los}$ & $v_{\rm turb}$ & $B_{\mparallel}$  & $|B_{\perp}|$ & $\phi_B$\\
\hline
   First & 4 & 1 & 0 & 1 & 1 & 1\\
   Second & 7 & 3 & 0 & 2 & 1 & 1\\
   Third & 9 & 7 & 4 & 3 & 2 & 1\\
\hline%\hline
\end{tabular}
\tablefoot{
The first cycle did not include the \ion{Ca}{ii}~K line. The microturbulence used in the first and second cycle was the given by the FAL-C model. Nodes are by default distributed equidistantly but in $|B_{\perp}|$ were placed at $\log(\tau)=[-1.0,-4.0]$.}
\end{table}

%%%%%%%%%%%%%%%%%%%%%%%%%%%%%%%%%%%%%%%%%%%%%%%%%%%%%%%%%%%%%%%%%%%%%%%%%
\begin{figure*}
\centering
\includegraphics[width=\linewidth]{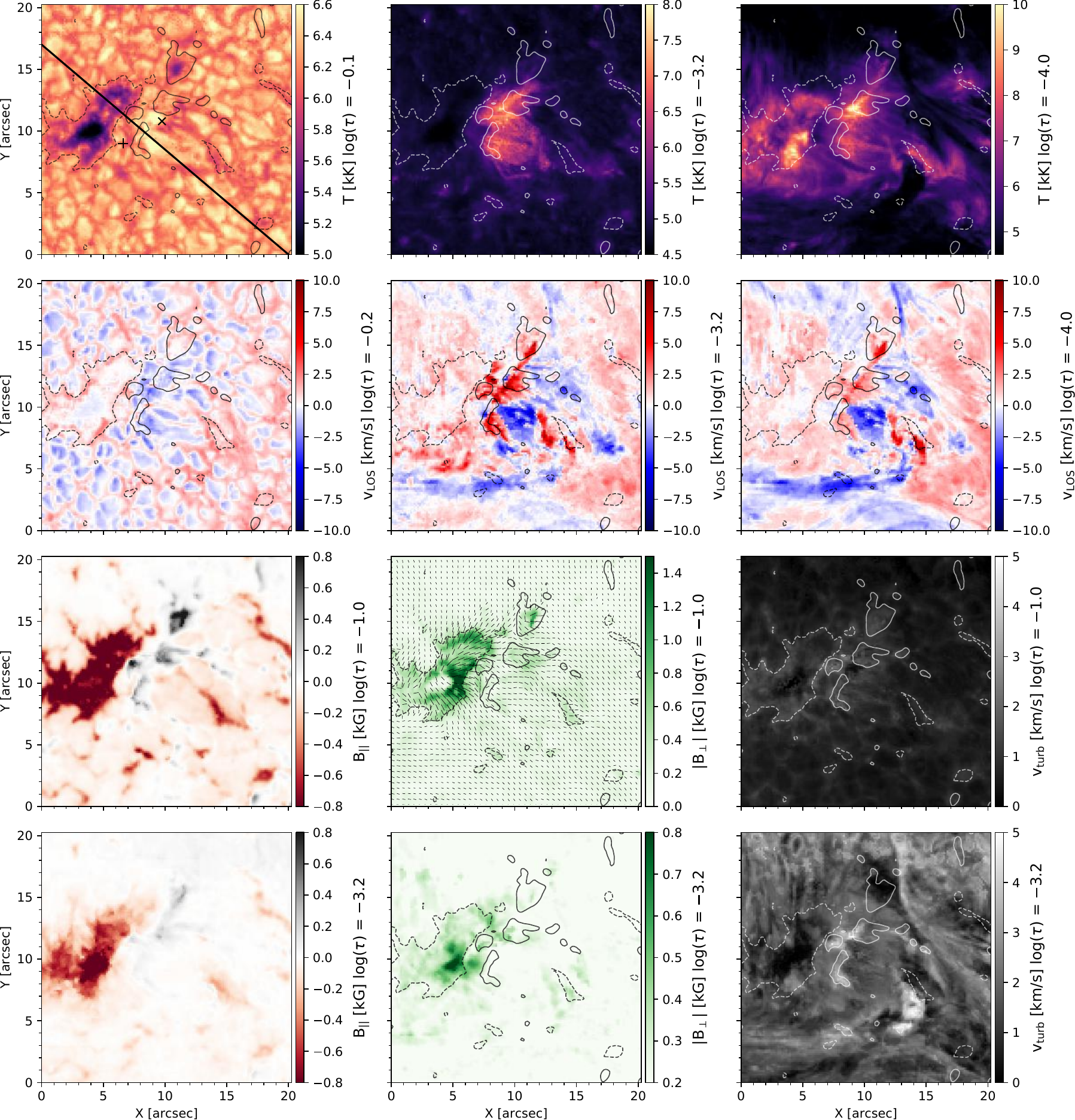}
\caption{Atmospheric structure in the ROI as inferred from the inversion. The first and second rows show the temperature and the line-of-sight velocity at three layers, third and fourth rows show the longitudinal magnetic field, the transverse magnetic field, and the microturbulent velocity at two layers. In the line-of-sight velocity maps, the blue color represents motions toward the observer (upflows) while the red color from the observer (downflows). Crosses in the upper-left panel show the location of the pixel examples of Fig.~\ref{fig:fullprofile} and the solid line the cross-cut shown in Fig.~\ref{fig:crosscut}. Thin dashed and dotted contours are contours of $B_{||}(\log (\tau)=0) = \pm 200$~G.}
\label{fig:inversion_maps}
% /scratch/carlos/CRISP/INVERSION/2PIXEL/SUPERPLOT
\end{figure*}

%%%%%%%%%%%%%%%%%%%%%%%%%%%%%%%%%%%%%%%%%%%%%%%%%%%%%%%%%%%%%%%%%%%%%%%%%
\begin{figure*}
\centering
\begin{subfigure}[t]{\textwidth}\centering
    \includegraphics[width=0.97\linewidth]{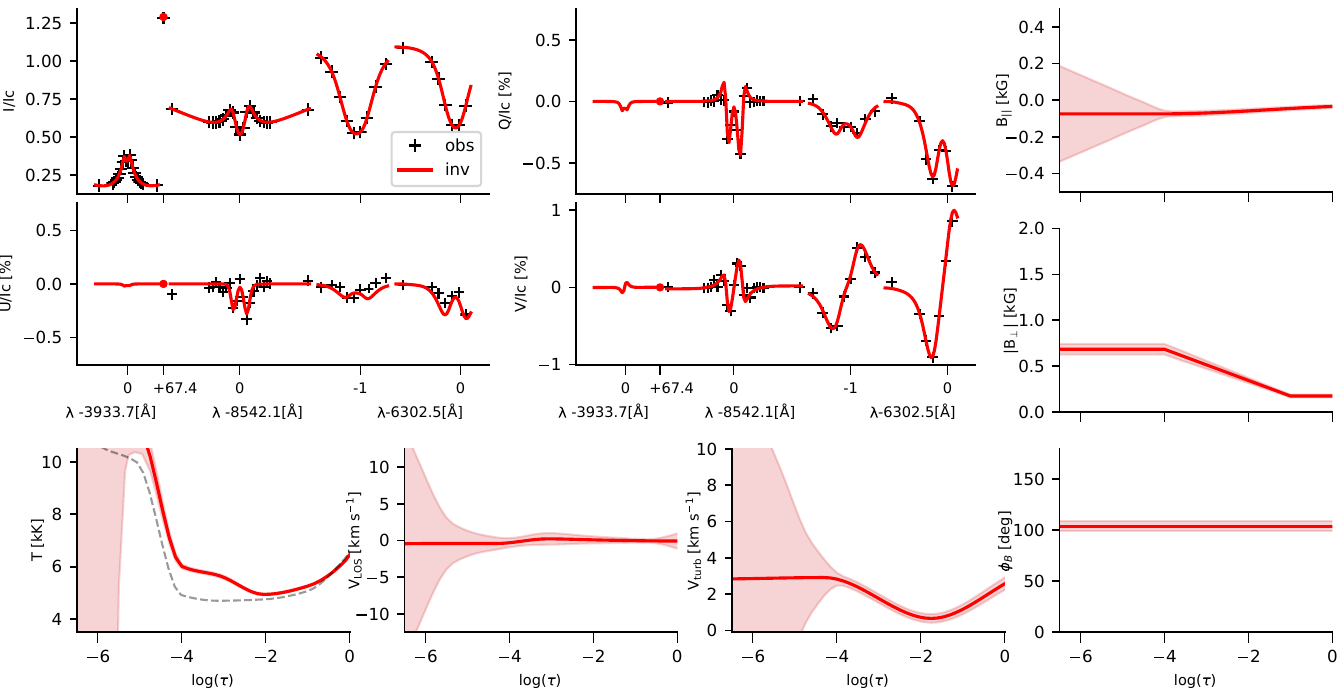}
    \caption{Pixel (6.6\arcsec, 9.0\arcsec) is shown as blue '+' symbol in Fig.~\ref{fig:inversion_maps}.} \label{fig:pixelQ}
\end{subfigure}\hfill%\vspace{0.025cm}
\begin{subfigure}[t]{\textwidth}\centering
    \includegraphics[width=0.97\linewidth]{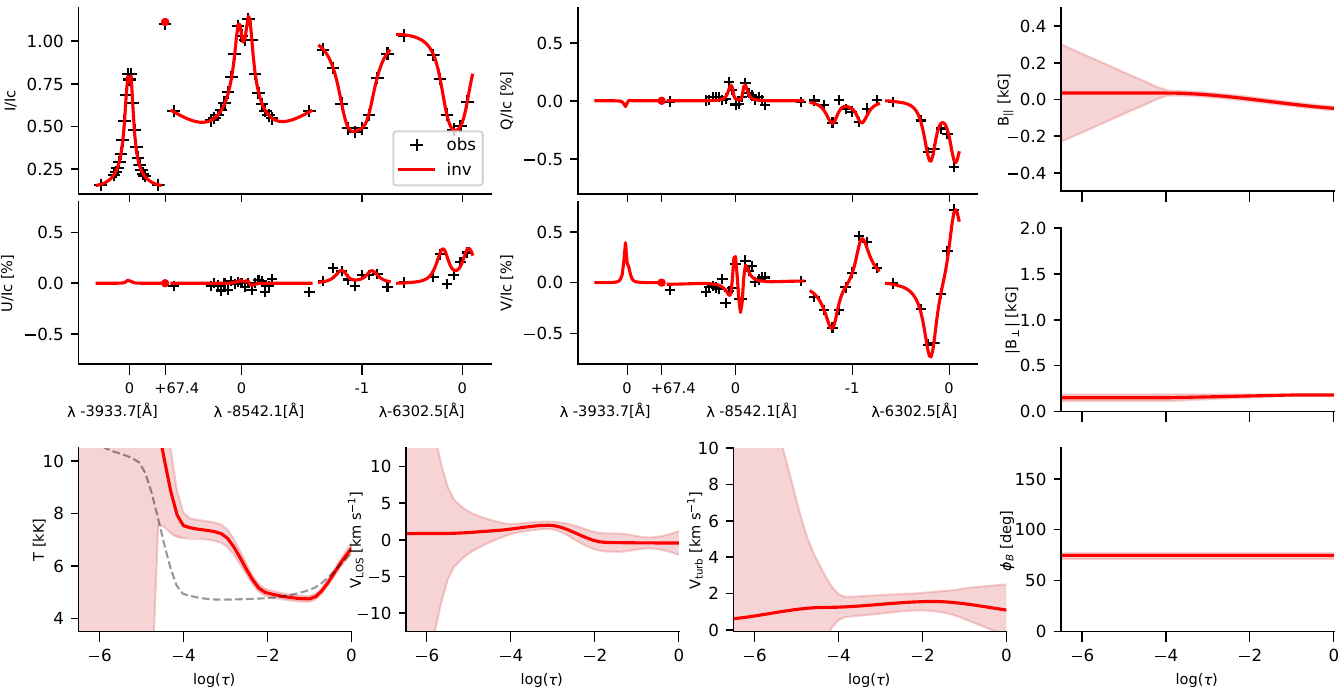}
    \caption{Pixel (9.7\arcsec, 10.8\arcsec) is shown as black 'x' symbol in Fig.~\ref{fig:inversion_maps}.} \label{fig:pixelF}
\end{subfigure}
\caption{Examples of spectra (observed in black and synthetic in red) with the corresponding inferred atmosphere. Red shaded bands indicate node uncertainties. The location of each pixel is quoted in each panel.   {The gray dashed curve represents the average quiet-Sun temperature stratification.}}
\label{fig:fullprofile}
% /scratch/carlos/CRISP/INVERSION/2PIXEL/
\end{figure*}

%%%%%%%%%%%%%%%%%%%%%%%%%%%%%%%%%%%%%%%%%%%%%%%%%%%%%%%%%%%%%%%%%%%%%%%%%
\section{Results}\label{sec:results}
%%%%%%%%%%%%%%%%%%%%%%%%%%%%%%%%%%%%%%%%%%%%%%%%%%%%%%%%%%%%%%%%%%%%%%%%%

%%%%%%%%%%%%%%%%%%%%%%%%%%%%%%%%%%%%%%%%%%%%%%%%%%%%%%%%%%%%%%%%%%%%%%%%%
\subsection{One snapshot inversion}\label{sec:snapshot}
%%%%%%%%%%%%%%%%%%%%%%%%%%%%%%%%%%%%%%%%%%%%%%%%%%%%%%%%%%%%%%%%%%%%%%%%%

The inversion of the whole FOV and/or the complete temporal series is very computationally expensive. Therefore, we have inverted our ROI only around 12:15\,UT. This snapshot includes the observation of the \ion{Fe}{i} lines (not included in the other two time-series), exhibits good seeing in all wavelength positions for all spectral lines, and the time difference between the observation of the central wavelength of \ion{Ca}{ii}~K and \ion{Ca}{ii}~8542\,\AA\ is less than 1 second. 
% Since the analysis is performed in an optical-depth axis, regions at equal $\tau$ values can sample different geometrical heights.
Since the analysis   {is shown with maps at a given optical depth}, regions at equal $\tau$ values can sample different geometrical heights.

%%%%%%%%%%%%%%%%%%%%%%%%%%%%%%%%%%%%%%%%%%%%%%%%%%%%%%%%%%%%%%%%%%%%%%%%%
\paragraph{Temperature stratification:} Figure~\ref{fig:inversion_maps} shows horizontal slices of the inferred temperature at three different optical depths: at $\log(\tau)\sim0$, associated with the continuum formation layer in the photosphere, $\log(\tau)=-3.2$ which shows the upper photosphere (around the solar temperature minimum in quiet-sun models), and at $\log(\tau)=-4$, which provides a view of the chromosphere. At $\log(\tau)\sim0$, the temperature shows the pattern of photospheric granulation with temperatures around 6000\,K. In this layer, there is no indication of chromospheric activity. At $\log(\tau)\sim-3.2$, we found a region of about $6\arcsec \times6\arcsec$ with a temperature of around 7000\,K compared with the $\sim$5000\,K of the surrounding. It starts at the edge of the pore where magnetic flux cancellation is expected to occur and extends towards the right side coinciding in location with the bright fibrils (see Fig.~\ref{fig:mosaico}). 

The atmosphere above (at $\log(\tau)=-4$) shows a very different temperature distribution. The region of the fibrils have now temperatures similar to what is found above the pore. In both places, some areas exceed 10\,kK. Outside these areas, temperatures are much lower. While the higher temperatures above the fibrils can be understood as a transfer of energy from below, the increase in temperature above the pore has no counterpart below and could then be generated at higher altitudes.

%%%%%%%%%%%%%%%%%%%%%%%%%%%%%%%%%%%%%%%%%%%%%%%%%%%%%%%%%%%%%%%%%%%%%%%%%
\paragraph{Velocity stratification:} The line-of-sight velocity maps are shown below the temperature panels in Fig.~\ref{fig:inversion_maps}. The velocity map at $\log(\tau)\sim0$ displays the typical granulation pattern, with upflows in the granules and downflows in the intergranular lanes. At $\log(\tau)=-3.2$ the low-chromospheric velocity map shows downflows and upflows with peaks of the order of $\pm$10\,\kms. Downward motions are found above the positive polarity whereas upward motions are right next to the downward motions. Since the plasma flows tend to align along magnetic field lines, the apparent upward and downward motions located next to each other suggest the presence of bidirectional flows along inclined magnetic field lines with respect to the line of sight.

There are also some downward motions near (14\arcsec, 7\arcsec) close the other small negative polarity, which could be hot plasma that is not able to escape and is redirected downwards again along low-lying field lines. Finally, one of the surges along y=3\arcsec\ appears to show upward motions. Higher in the atmosphere at $\log(\tau)=-4.0$ the general trend is similar;  the upflows persist while the downflows are less intense.

% 
% The presence of bidirectional flows next to each other suggests flows along magnetic field lines with a certain inclination angle relative to the line of sight, assuming the flows parallel to the field.

%%%%%%%%%%%%%%%%%%%%%%%%%%%%%%%%%%%%%%%%%%%%%%%%%%%%%%%%%%%%%%%%%%%%%%%%%
\paragraph{Magnetic field stratification:} The inferred components of the magnetic field relative to the line-of-sight ($B_{\mparallel}$ and $|B_{\perp}|$) are also presented in Fig.~\ref{fig:inversion_maps}.

The longitudinal magnetic field $B_{\mparallel}$ in the photosphere at $\log(\tau)=-1.0$ shows the pre-existing pore (at the left side of the ROI) with a longitudinal magnetic field larger than $-1000$\,G. The emergence of the magnetic bubble brings to the surface a positive polarity patch that interacts with the pore and a negative polarity concentration visible in the lower-right part of the region (14\arcsec, 4\arcsec). In the chromosphere at $\log(\tau)=-3.2$, the magnetic field is more diffuse and weaker although maintaining the general topology of the photospheric field. While in the pore the longitudinal magnetic field decreases to 800\,G in the chromosphere, the longitudinal magnetic field in the location of the bright fibrils is very weak.\\

The transverse magnetic field $|B_{\perp}|$ in the photosphere at $\log(\tau)=-1.0$ correlates with the areas of a strong longitudinal magnetic field (highlighted in this panel with contours at $\pm200$\,G). The transverse field reaches more than 1.5\,kG in the right edge of the pore, and decreases rapidly outside the latter. The direction of the inferred azimuth ($\phi_B$) has also been drawn on this panel. Although it has not been corrected from its 180-degree ambiguity, it is enough to recognize that the general direction of the field is radial to the interior of the pore. In the chromosphere at $\log(\tau)=-3.2$, the transverse magnetic field is in general much weaker, except in the region between the pore and the new polarity around (7\arcsec, 10\arcsec) where the field increases up to 700\,G compared to the very weak photospheric field at this location. At the location of the fibrils, the transverse magnetic field is of the order of 200\,G. Given that the longitudinal component is around 120\,G, the inclination of the field lines would be around 30 degrees. However, given the low field strength and the noise present in the signals, we cannot ensure the reliability of this estimation.

%%%%%%%%%%%%%%%%%%%%%%%%%%%%%%%%%%%%%%%%%%%%%%%%%%%%%%%%%%%%%%%%%%%%%%%%%
\paragraph{Microturbulent velocity:} In some cases, STiC is not capable of finding a temperature stratification that allows explaining the width of these profiles without including a microturbulent contribution to the Doppler width. In the photosphere, the microturbulence has values below 1~\kms\ on average, reaching $\sim$2~\kms\ in the intergranules and near the boundaries of the pore. This has been associated with turbulent motions in the intergranular lanes \citep{Borrero2002}. In the chromosphere (between $\log(\tau)=-3$ and $\log(\tau)=-4$) the microturbulent velocity increases on average up to 2-3~\kms\ in places far from the cancellation site. In locations with strong downflows, it reaches higher values (3-4~\kms). This increase could be due to the turbulent motions as the plasma is being compressed towards the denser lower atmosphere. On the other hand, we found that the plasma associated with upward motions does not require such additional broadening (see Fig.~\ref{fig:inversion_maps}). This might indicate how the plasma can escape easier to higher layers of the atmosphere where the surrounding density is lower.

%%%%%%%%%%%%%%%%%%%%%%%%%%%%%%%%%%%%%%%%%%%%%%%%%%%%%%%%%%%%%%%%%%%%%%%%%
\paragraph{Spectral signatures:} 
To show the shape of the typical observed Stokes profiles and the quality of the fits, some profiles have been extracted from the location marked in the temperature map at $\log(\tau)\sim0$ in Fig.~\ref{fig:inversion_maps} and shown in Fig.~\ref{fig:fullprofile}. Overall, the fits to the Stokes profiles follow the observations closely. The first pixel~\ref{fig:pixelQ} was extracted from the region of the high transverse chromospheric magnetic field. In this location, the wings of the \ion{Ca}{ii} lines are in emission while the core is in absorption. Although weak, the \cair signals are above the noise and have been reproduced by the inversion code.

One more profile was extracted from the bright fibrils and it is shown in Fig.~\ref{fig:pixelF}. The line-core intensities of \ion{Ca}{ii} 8542\,\AA\ and \ion{Ca}{ii}\,K are high, up to 13 and 3 times compared to their typical values at the quiet Sun, even higher than during some strong flares \citep{Kuridze2018, Rahul2020}. The retrieved atmosphere shows a strong temperature enhancement in the chromosphere compared to the profile in panel \ref{fig:pixelQ}. One consequence of that is the loss of sensitivity higher up as the material ionizes. This behavior can be detected in the uncertainties calculated for this pixel compared with the previous pixel. We note that our estimate of the uncertainties does not include degeneracies between different parameters and it does not allow to estimate errors included in the assumptions of the models such as the 1.5D plane-parallel approximation, the assumption of hydrostatic equilibrium, etc. Our estimate for one parameter is purely based on the goodness of the fit and the sensitivity of the different spectral lines in all Stokes parameters to a few nodes in the atmosphere.

%We note that the calculated uncertainties might be underestimated (as each physical parameter is considered independent) and the uncertainties of quantities with very few nodes (like the magnetic parameters) represent only the range of configurations compatible with the profiles under such reduced flexibility and not the response of the profiles at every point of the atmosphere. 

%%%%%%%%%%%%%%%%%%%%%%%%%%%%%%%%%%%%%%%%%%%%%%%%%%%%%%%%%%%%%%%%%%%%%%%%%
\subsection{Evolution of the polarization signals in time}\label{sec:pol_signals}
%%%%%%%%%%%%%%%%%%%%%%%%%%%%%%%%%%%%%%%%%%%%%%%%%%%%%%%%%%%%%%%%%%%%%%%%%

\begin{figure}
\centering
\includegraphics[width=1.0\linewidth]{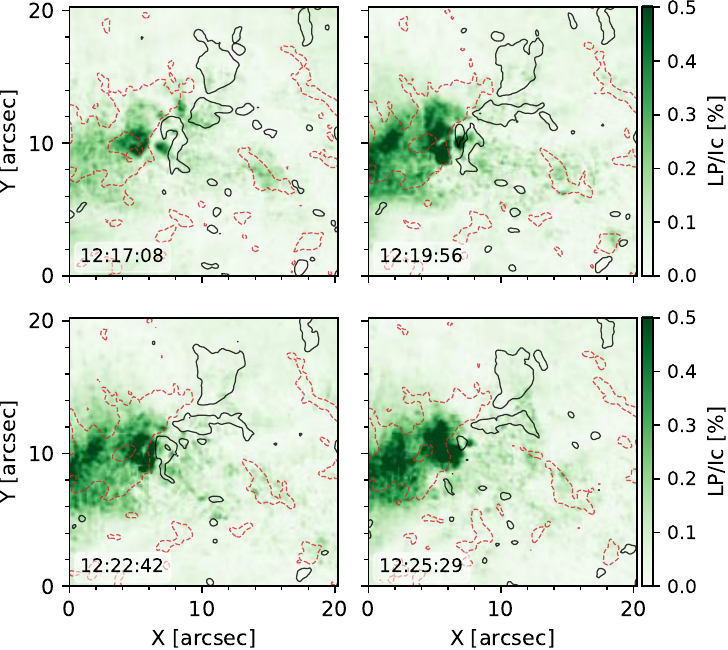}
\caption{Linear polarization maps at the central wavelength of the \cair line. The polarity is highlighted with contours at $-3\cdot10^{-3}I_c$ and $5\cdot10^{-4}I_c$ of Stokes $V$ in the red wing of the \ion{Fe}{i} 6301\,\AA\ line. The time of the observations is shown in the lower-left corner of each panel.}
\label{fig:polmap}
% /scratch/carlos/CRISP/INVERSION/2PIXEL/1temporalEvolution_paper// spython plotMagneticPAPER_v3.py
\end{figure}  

Although we do not discuss in detail the overall temporal evolution of our target, we briefly discuss the evolution of the polarization signals. Figure~\ref{fig:polmap} shows the evolution of the linear polarization map of the \cair line, and therefore the evolution of the chromospheric magnetic field indirectly. Some contours show the temporal evolution of the circular polarization signals in the photosphere (given by the \ion{Fe}{i} 6301\,\AA\ line). The first panel resembles the transverse magnetic field map shown in Fig.~\ref{fig:inversion_maps}, indicating that the inversion code was able to extract this chromospheric information. 

Two main conclusions can be drawn from Fig.~\ref{fig:polmap}: the circular polarization is decreasing as the magnetic flux is canceled, while the linear polarization increases in amplitude and area. This enhanced linear polarization is consistent with a $\cap$-configuration as the field lines are being bent or reconfigured during the magnetic reconnection. These polarization signals indicate a transverse magnetic field of 1\,kG in a large part of the region. These unusually strong transverse fields (more than 1\,kG) have been also found in the chromosphere during magnetic reconnection events \citep{Kuridze2018}.

\begin{figure}
\centering
% \hrule
\includegraphics[width=\linewidth,trim={0 33pt 0 0},clip]{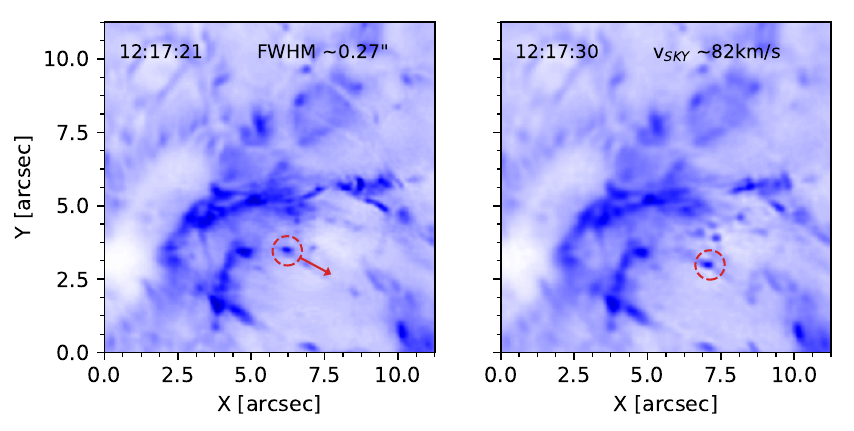}
\includegraphics[width=\linewidth,trim={0 33pt 0 0},clip]{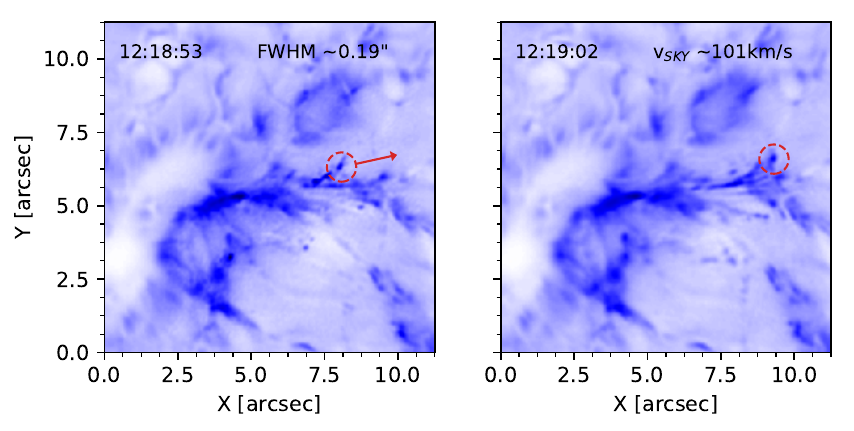}
\includegraphics[width=\linewidth,trim={0 5pt 0 0},clip]{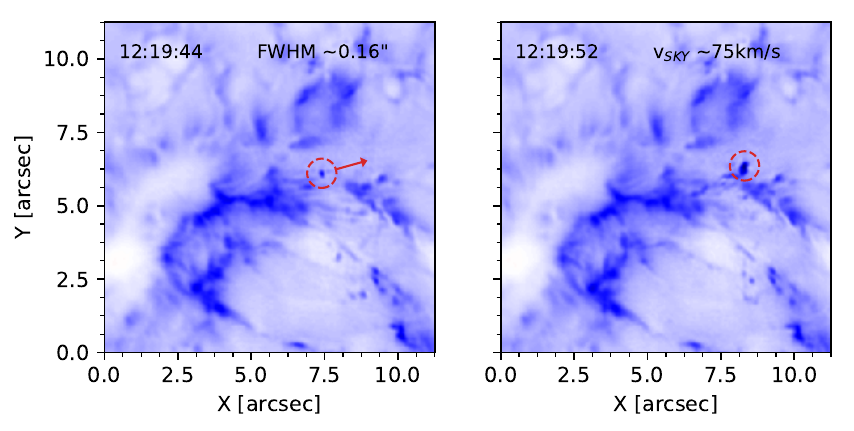}
\caption{Monochromatic images in the blue wing of \ion{Ca}{ii} K ($\lambda_0-1.3\AA$) at three temporal intervals. Each row shows a different blob. Red circles and arrows indicates the location and direction of the motion. FWHM inferred from the blobs are quoted in the right corner of the right panels. Transverse velocities are shown in the left panels.}
\label{fig:jets_vsky}
% /scratch/carlos/CRISP/JETS/calculatev_JUNTOS.py
\end{figure}

%%%%%%%%%%%%%%%%%%%%%%%%%%%%%%%%%%%%%%%%%%%%%%%%%%%%%%%%%%%%%%%%%%%%%%
\subsection{Plasmoid-mediated magnetic reconnection}\label{sec:plasmoid}

During the evolution of this event, we have detected the presence of many small-scale blobs with similar shapes and widths, between 0.15 and 0.3\arcsec, moving away from the cancellation site, identified in the blue wing ($\lambda_0-1.3\AA$) of the \ion{Ca}{ii}~K line. These blobs have been found previously in the solar atmosphere \citep{Innes2015, Rouppe2017}, and have been identified as plasmoids. They are formed when the current sheet exceeds a certain aspect ratio, becomes unstable, and fragments into smaller elements (plasmoids). If the latter are detected during at least two times steps we can track them and infer their apparent speed in the plane of the sky. Figure~\ref{fig:jets_vsky} shows three examples of these blobs at different time frames with the inferred velocities. We detect fast transverse velocities between 70\,km/s and 100\,km/s. Their high velocities allow us to distinguish them easier from the surroundings at this wavelength where they are clearly visible over a more photospheric background. Given that both the LOS and plane-of-the-sky velocities are of the same order, the ejected plasma appears to be neither predominantly vertical nor horizontal, supporting the idea of a tilted current sheet assuming that the flows are parallel to that. The direction of the motions is pointing away from the main polarity (pore) that the small opposite polarity is moving into. This is consistent with other observations   {and} simulations \citep{Chitta2017, Peter2019} as the stronger vertical magnetic field of the dominant polarity might be more difficult to bend. We identified more than 25 similar blobs in the three time-series (detected in two consecutive frames). We also have hints of other blobs, but they are more difficult to identify as they last less than 9 seconds (one frame). This means that on average we have more than 1 blob every 30 seconds. This value is compatible with the timescale of plasmoid formation found in simulations \citep{Nobrega2016}. The presence of plasmoids further supports the magnetic reconnection scenario.

% in the black-dashed square shown in Fig.~\ref{fig:mosaico}

\begin{figure}
\centering
\includegraphics[width=\linewidth]{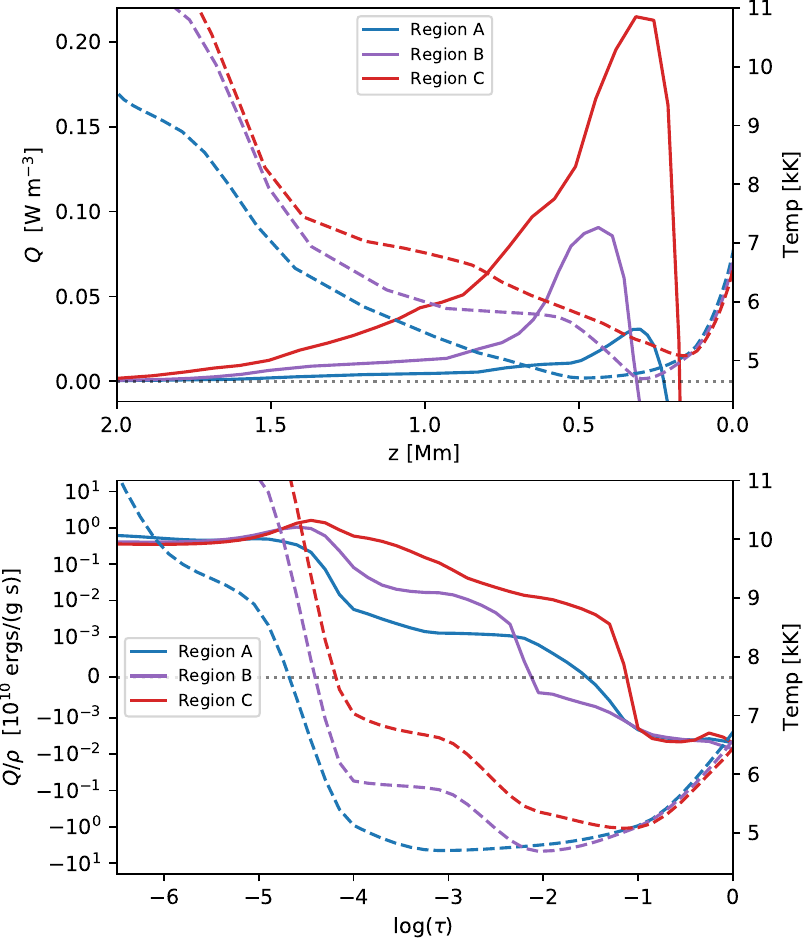}
\caption{Height distribution of the total radiative cooling rate per unit volume $Q$ (upper panel) and per unit mass $Q/\rho$ (lower panel) in solid lines calculated as the average of the regions highlighted in Fig.~\ref{fig:cooling} with their average temperature stratification (dashed lines).}
\label{fig:heightcooling}
% /scratch/carlos/stic_RadiativeLosses
\end{figure}

% \clearpage
%%%%%%%%%%%%%%%%%%%%%%%%%%%%%%%%%%%%%%%%%%%%%%%%%%%%%%%%%%%%%%%%%%%%%%%%%
\subsection{Energy balance}
%%%%%%%%%%%%%%%%%%%%%%%%%%%%%%%%%%%%%%%%%%%%%%%%%%%%%%%%%%%%%%%%%%%%%%%%%

We can estimate the radiative cooling rate per unit volume from our inferred models as the frequency-integrated radiative flux divergence: $ \int \nabla F_\nu \mathrm{d} \nu$. We approximate this by summing the net cooling rates in bound-bound transitions for the spectral lines in our model atoms for \CaII\ and \MgII:
\begin{equation}
    Q = h \nu_0 (n_u R_{ul} - n_l R_{lu}),
\end{equation}
where $h$ is the Planck's constant, $\nu_0$ the line center frequency, $n_u$ and $n_l$ the population of each level, and $R_{ul}$ the radiative rate coefficient from level $u$ to level $l$.
This is an approximation: in the photosphere we ignore contributions from many weak spectral lines and the H$^-$ continuum, in the upper chromosphere we ignore Ly$\alpha$ and the transitions of helium. In the low and mid-chromosphere, this approximation is expected to be rather accurate: the heating in the Balmer continuum and the cooling in the H$\alpha$ line tend to cancel each other, and other transitions tend to contribute much less to the cooling than \CaII\ and \MgII\ \citep{Vernazza1981}.

\begin{figure}
\centering
\includegraphics[width=\linewidth]{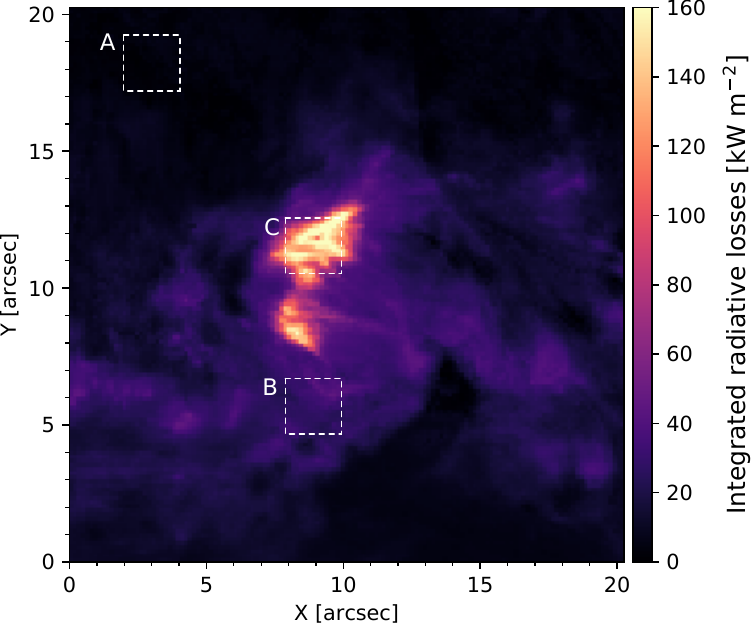}
\caption{Total integrated radiative losses calculated from the semi-empirical model inferred in the inversion.}
\label{fig:cooling}
% /scratch/carlos/stic_RadiativeLosses/fullexample
\end{figure}

The height distribution of the total radiative cooling rate per unit volume ($Q$) and per unit mass ($Q/\rho$) is shown in Fig.~\ref{fig:heightcooling} for three different patches inside the ROI.  We chose a quiet-sun region (A), a region close to the fibrils (B), and a region inside the brightest fibrils (C), all indicated in Fig.~\ref{fig:cooling}. As expected, higher temperature regions in the atmosphere correlate with enhanced radiative cooling. The radiative losses are largest at the bottom of the chromosphere because the density is also much larger deeper down.

\begin{table}
\caption{Average contribution of each spectral line to the total integrated radiative losses for the regions highlighted in Fig.~\ref{fig:cooling}, in units of kW~m$^{-2}$.}
\label{tab:cooling}
\centering
\begin{tabular}{c c c c c c c c c}
\hline\hline
\small Region/Line & \small Ca K &  \small Ca H & \small Ca IR$^{\dagger}$ & \small Mg k & \small Mg h &\small Total \\
\hline
   A & 0.8 & 0.7 & 2.6 & 0.1 & 0.3 & 4.5\\
   B & 7.6 & 5.6 & 12 & 3.7 & 3.1 & 32\\
   C & 28 & 20 & 30 & 16 & 14 & 108\\
\hline%\hline
%   Total & 9 & 7 & 4 \\
% \hline
\end{tabular}
\tablefoot{ ${\dagger}$ Ca IR refers to the total of the three lines of the \ion{Ca}{ii} infrared triplet at 8542\,\AA, 8498\,\AA, 8662\,\AA.}
\end{table}
%%%%%%%%%%%%%%%%%%%%%%%%%%%%%%%%%%%%%%%%%%%%%%%%%%%%

The total radiative losses in each pixel have been integrated over geometrical height (which is derived assuming hydrostatic equilibrium in the inversion) between the height of the temperature minimum and the height where the temperature reaches 10\,kK. Above that temperature, the \ion{Ca}{ii} lines lose sensitivity and the hydrogen Ly$\alpha$ line starts to contribute, so our derived cooling rates are inaccurate there. The total integrated radiative losses for all the pixels of the FOV are shown in Fig.~\ref{fig:cooling}, and the contribution of each spectral line to these losses is given in Table~\ref{tab:cooling}.

In the quiet sun (region A) we obtain around $4.5$~kW~m$^{-2}$, while in the intermediate area (region B) we obtain losses of $32$~kW~m$^{-2}$. These values agree well with the previous estimates \citep{Vernazza1981, Withbroe1977}. We emphasize that our cooling rate estimates are more consistent with the real situation: they are computed for each spatially resolved atmosphere and then averaged, whereas previous studies first averaged observed line profiles, then inferred an atmosphere consistent with the average line profile, and finally computed the cooling rate from that atmosphere. 

In region C the average integrated radiative losses are $108$~kW~m$^{-2}$ but Fig.~\ref{fig:cooling} shows peak values up to 160~kW~m$^{-2}$. These values are roughly a factor 5 higher than the semi-empirical values for active regions, indicating that the averaging of the observations can lead to a substantial underestimation of the local radiative cooling rate.

Finally, we show the correlation of the height-integrated radiative losses with the wavelength-integrated intensity in the \ion{Ca}{ii}~K line. The wavelength integration was done between the K$_1$ minima whose formation is close to the temperature minimum \citep{Bjorgen2018}. The correlation is strong and close to linear, with a Pearson coefficient of 0.97. This result supports the analysis and results of \citet{Leenaarts2018}, who used wavelength-integrated \ion{Ca}{ii}~K intensity as a proxy for chromospheric radiative losses.

\begin{figure}
\centering
\includegraphics[width=\linewidth]{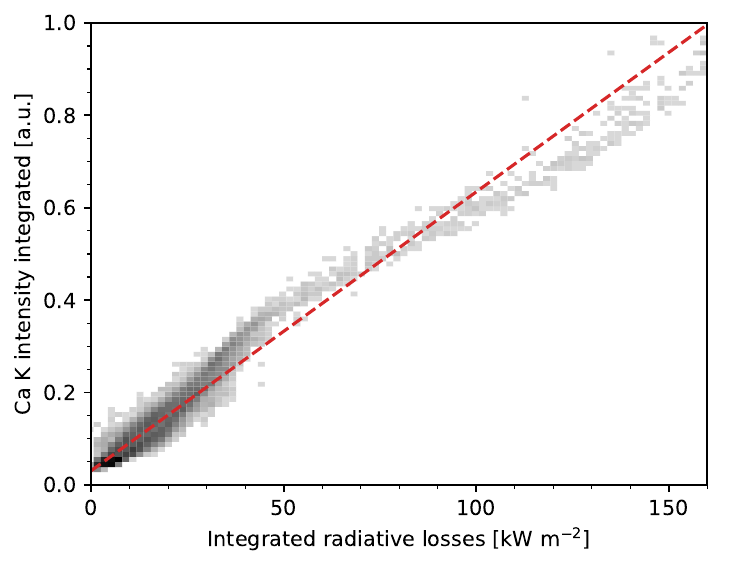}
\caption{Joint probability distributions of the wavelength-summed \ion{Ca}{ii} K intensity and the total integrated radiative losses computed from the data shown in Fig.~\ref{fig:cooling}. The color of each bin is proportional to the logarithm of the number of points, with darker color indicating more points.}
\label{fig:corr}
\end{figure}

As the energy lost during the whole observation needs to be continually replenished, the total integrated radiative losses in the ROI assuming a similar emission during the two hours is around $\sim5\cdot10^{27}$\,erg. Motivated by the magnetic nature of this event, we propose that the energy required for this event can be extracted from magnetic reconnection in the chromosphere. 
We can use the magnetic field vector from the inversion to estimate the magnetic energy content of the atmosphere, by integrating the whole volume $1/8\pi \int B^2 dV$ including the stratification in height. In that case, the energy content of this volume is $\sim5\cdot10^{29}$\,erg. We have set the integration range from $\log(\tau)=0$ to $\log(\tau)=-4$ which corresponds approximately to a height range of $\Delta z=800$\,km where we have enough magnetic sensitivity. 

% horizontal component and the
%This value is around 1 order of magnitude larger than in the calculated from SDO.

%\fixed{The magnetic energy converted during the flux cancellation can be estimated as $1/8\pi B^2 V = 1/8\pi\Phi^2H/A$, where inserting the previously estimated total flux of the pore $\Phi\sim 4\cdot10^{19}$~Mx, assuming that the large part of the energy content is below $H\sim500$~km as decreases rapidly with height, with an estimated size of $A\sim  5\times5$ arcsec$^2$, we obtain a value of $\sim5\cdot10^{28}$\,erg which is larger than the radiative losses and then sufficient to inject that energy into the chromosphere.}
% (A = 3750km *3750km, H = 500 km, FLUX = 4e19 Mx).
% $A\sim  1.4\cdot10^{7}$~km$^{2}$

% /scratch/carlos/stic_RadiativeLosses/fullexample

% In our intermediate temperature area (region B) we obtain losses of $32$~kW~m$^{-2}$, once again consistent  (if somewhat higher) with earlier semi-empirical work based on spatially-averaged observations \citep{Vernazza1981, Withbroe1977}. 

%%%%%%%%%%%%%%%%%%%%%%%%%%%%%%%%%%%%%%%%%%%%%%%%%%%%%%%%%%%%%%%%%%%%%%%%%
%%%%%%%%%%%%%%%%%%%%%%%%%%%%%%%%%%%%%%%%%%%%%%%%%%%%%%%%%%%%%%%%%%%%%%%%%

\begin{figure}
\centering
\includegraphics[width=\linewidth]{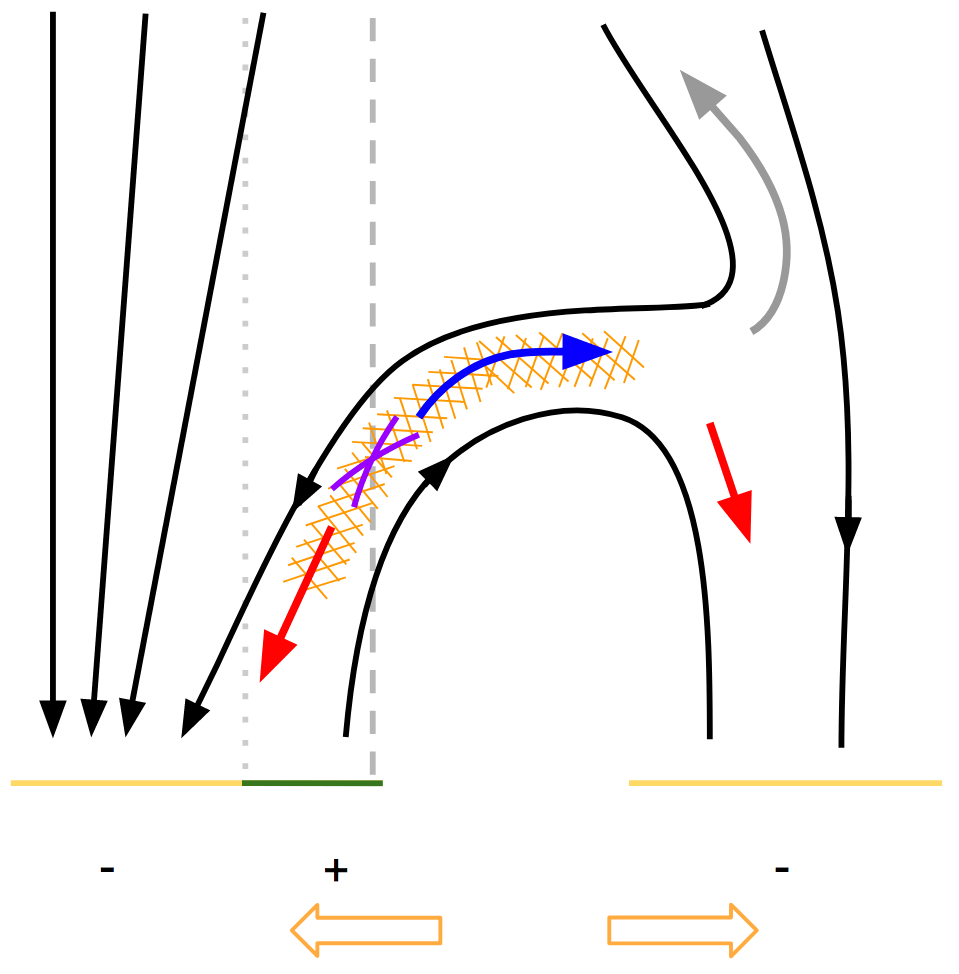}
\caption{Schematic illustration  of  the  proposed  magnetic  field  topology. Suggested magnetic field lines are given in black. Upflows are colored in blue while downflows in red. Vertical dotted and dashed lines are drawn for comparison with Fig.~\ref{fig:crosscut}.}
\label{fig:cartoon2}
\end{figure}

\begin{figure}
\centering
\includegraphics[width=\linewidth]{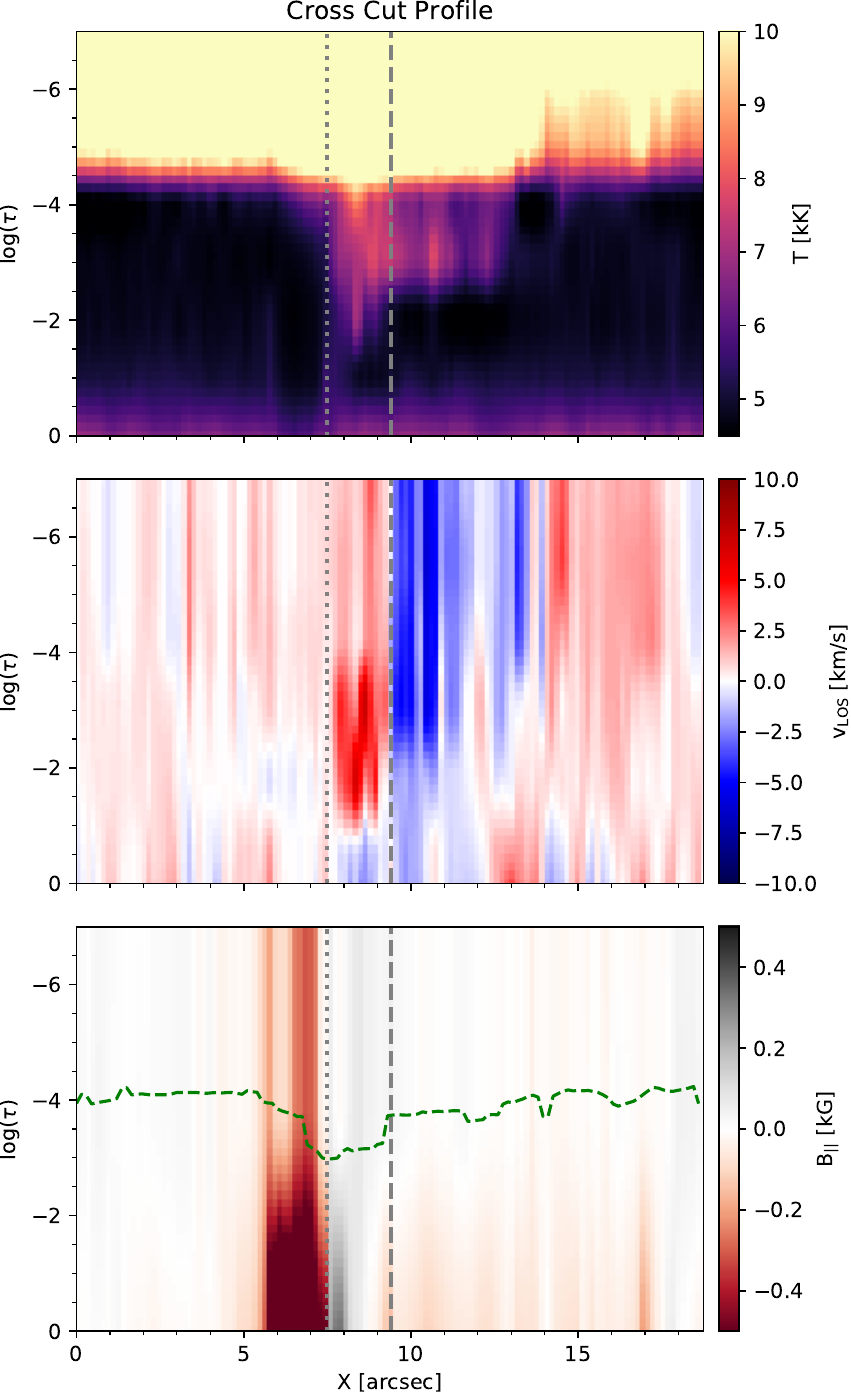}
\caption{Vertical stratification across the solid line indicated in the upper-left panel of Fig.~\ref{fig:inversion_maps}. From top to bottom: temperature, line-of-sight velocity and longitudinal magnetic field.}
% /scratch/carlos/CRISP/INVERSION/2PIXEL/RFBT
\label{fig:crosscut}
\end{figure}

%%%%%%%%%%%%%%%%%%%%%%%%%%%%%%%%%%%%%%%%%%%%%%%%%%%%%%%%%%%%%%%%%%%%%%%%%
\subsection{Magnetic field topology}
%%%%%%%%%%%%%%%%%%%%%%%%%%%%%%%%%%%%%%%%%%%%%%%%%%%%%%%%%%%%%%%%%%%%%%%%%
% \clearpage spatial decoupling
The analysis of the monochromatic images and the inversion results have revealed several aspects that can be put together to propose a magnetic topology consistent with the following findings:

\begin{enumerate}
    
    \item The emergence of a new small magnetic flux concentration causes the footpoints of opposite polarity to separate and the granulation is deformed accordingly. It indicates that the flux tube adopts the shape of an $\Omega$ loop.
    
    \item The positive polarity of a newly emerged magnetic concentration interacts with a pre-existing pore of a fully-developed active region.
    
    \item In the location where these two polarities come into contact and cancellation of the magnetic flux occurs, a strong emission is found in all chromospheric/coronal diagnostics.  
    
    \item The temperature inferred by the inversion code indicates that in the lower chromosphere and above the temperature increases for more than 2000~K while no signs of activity are present in the photosphere.
    
    \item Bidirectional jets (accelerated to $\pm$ 10\kms) are found with downward motions located above the positive polarity and upward motions towards the negative polarities. This indicates that the plasma flows are very inclined with respect to the line of sight. The apparent motion in the plane of the sky (detected in the monochromatic images) is also pointing away from the pore.
    
    \item After reaching a certain height, given the topology of the field, the plasma above seems to move towards the left side of the FOV as the surges indicate. According to Fig.~\ref{fig:sdomaps}, the magnetic field at higher layers might be connected to the sunspot of the left side of the active region.

\end{enumerate}

All the above findings are suggestive of a model magnetic reconnection between an emerging magnetic field and the pre-existing field, as found in observations as well as simulations \citep{Nobrega2016, Hansteen2019, Guglielmino2019}. The topology of our event is expected to be very similar to the classical emerging flux model for solar reconnection \citep{Heyvaerts1977, Yokoyama1995} but on a much smaller scale where the event occurs deeper in the lower chromosphere. In this scenario, a magnetic diffusion region would arise around the interface between the two magnetic polarities. Inside this diffusion region, free magnetic energy is released in several different ways: the electric field induced by the magnetic field creates a thin layer of an intense current   {sheet}, which heats the local plasma through Joule heating. In addition, plasma is ejected away from the reconnection site by the magnetic tension force. Another difference with those models is that the magnetic field at higher layers in many of the simulated cases is nearly vertical, while here the numerous plasma flows are in the plane of the sky and elongated structures indicate a nearly horizontal canopy of the pre-existent magnetic field.

Figure~\ref{fig:cartoon2} shows a simplified illustration of the proposed magnetic topology that is consistent with our results. In this figure we show schematically: a) the emerged magnetic loop and the motion of the footpoints (with orange arrows), b) the pore with strong vertical fields (negative concentration on the left side), c) the reconnection region (with a purple cross) with the corresponding heating (orange grid) and bidirectional flows (red and blue arrows) and d) later plasma motions which escape from the region (gray arrow) or which fall again (right red arrow). To compare the similarities of our 2D sketch, we also show in Fig.~\ref{fig:crosscut} a cross-cut of the inversion results across the polarities. In these figures, some vertical lines highlight the different domains and their correspondence in both figures. In this figure, the extended heating occurring at $\log(\tau)=-3.5$ is clearly visible, similarly to \citet{Leenaarts2018}, which is located deeper in regions with downflows. The transition region is reconstructed at higher optical depth in the region showing temperature enhancements. This illustration also helps to understand how the reconnection site (dashed line) could take place to the west side of the cancellation site (dotted line) due to a geometrical perspective. We also added to the bottom panel of Fig.~\ref{fig:crosscut} the location of maximum sensitivity of the linear polarization to the transverse magnetic field in the \cair\ line. On average, it is placed around $\log(\tau)=-4$ in quiet-sun areas and $\log(\tau)=-3$ in the flaring region due to the temperature enhancement, as shown in \citet{Rahul2020}.

The dark jets observed in \Halpha\ at high velocity seems to match with the description of surges as the cool chromospheric jets that coexist with the hot jets \citep{Takasao2013, Nobrega2016}. It has been shown that the simple reconnection as a driver seems though insufficient to explain their higher altitude (barely detected in the \ion{Ca}{ii} lines), longitude (compared with the size of the hot fibrils), and velocities (compared with the reconnection outflow velocity). The acceleration mechanism found in numerical simulations has been shocks \citep{Takasao2013, Shibata2007}. The magnetic reconnection happening in the lower atmosphere can generate slow mode waves. These waves propagate into regions of lower density where eventually they form a shock as a result of the law of conservation of energy which accelerates the local plasma upwards \citep{Takasao2013}. These waves can be also generated when reconnection outflows hit the ambient magnetic field (right side of Fig.~\ref{fig:cartoon2}). Given the large time scale of our event, a recurrent generation of waves seems to match with our scenario, already suggested by \cite{Takasao2013}. To prove the wave/shocks scenario as the origin, we would need a better cadence in the observations.

  {Given the dynamic nature and spectral signatures of this observation, an alternative explanation could be given by a flaring atmosphere scenario. In the standard flare model, the energy released in the corona can be transported and dissipated into the chromosphere and even the photosphere. This scenario seems unlikely because a) the new small magnetic flux emergence should start interacting with the pre-existing canopy of the pore from the bottom of the atmosphere, b) the shape of the light-curves do not show a sudden peak and gradual decrease like in standard flares but rather a gradual evolution during the cancellation period and, c) it would be a very small flare (as it is not detected in GOES) and small flares are unlikely to produce the deep-seated and persistent heating for almost two hours.}

% \clearpage
%%%%%%%%%%%%%%%%%%%%%%%%%%%%%%%%%%%%%%%%%%%%%%%%%%%%%%%%%%%%%%%%%%%%%%%%%
\section{Summary and conclusions}\label{sec:conclusions}
%%%%%%%%%%%%%%%%%%%%%%%%%%%%%%%%%%%%%%%%%%%%%%%%%%%%%%%%%%%%%%%%%%%%%%%%%

Using multi-line non-LTE inversions, we retrieved a 3D atmospheric model of a reconnection event caused by the emergence of new magnetic flux into a previously-existing magnetic field configuration in the chromosphere. The addition of four lines   {respect to single-line inversions} and in particular the \ion{Ca}{ii}~K, has allowed us to improve the reconstruction of the atmospheric parameters to higher heights, while better constraining the areas of the model where the sensitivity of several lines overlap. Based on the results of the inversion and the appearance of the observed target in H$\alpha$, we propose a likely magnetic field configuration that is compatible with all observational signatures and with the inversion results. In the region where the reconnection occurs, we infer a temperature increase in the low chromosphere that is about 2000~K higher than in the surrounding quiet Sun, whereas higher-up in the chromosphere we find a 1000-3000~K increase in temperature in a more extended region surrounding the reconnection event. In both cases, we find a fine spatial structure down to sub-arcsecond scales. We also found bidirectional flows of the order of 10\,\kms\ emanating from the region as well as bright blobs of roughly 0.2~arcsec diameter in the \ion{Ca}{ii}\,K ejected from the same region at 100\,\kms.

Increased brightness in the 304\,\AA\ and 171\,\AA\ channels of SDO/AIA indicate heating associated with the reconnection event to transition region and coronal temperature. Our inversion is not sensitive to temperatures above approximately 10,000~K so we cannot set constraints on the height where the heating observed in AIA is located. A coronal response is not normally seen in photospheric/low-chromospheric reconnection events such as Ellerman bombs, but heating to transition-region temperatures in reconnection events higher in the chromosphere can give a response in AIA  if the column density of overlying cool material is low enough so that not all radiation is absorbed in the hydrogen and helium continua \citep{Young2018, Guglielmino2019, Hansteen2019, Silva2020}.

We have estimated the chromospheric radiative losses from the inferred atmospheric model in the reconnection target. The values that we obtain for the quiet Sun surrounding the active region are of the order of $\sim4.5$\,kW\,m$^{-2}$, consistent with earlier estimates based on semi-empirical models of the quiet Sun \citep{Vernazza1981, Withbroe1977}. In the active region, but outside of the reconnection event, we find values of 20--50\,kW\,m$^{-2}$, which is comparable with earlier modeling, but we also note substantial variation with location on the Sun at scales of an arcsecond. Inside the reconnection event, we find peak values of the radiative losses up to 160\,kW\,m$^{-2}$, with spatial variations on sub-arcsecond scales. These losses are a factor five or more than the typical value for the chromosphere in active regions.

The majority of the radiative losses occur at relatively low heights, around $300-500$~km above the $\log(\tau)=0$ surface, caused by a deep-seated chromospheric temperature rise. Whether these losses are counterbalanced by in-situ conversion from magnetic energy, or caused by relatively hot flows coming from larger heights is still an open question. The magnetic energy release estimated from the flux cancellation ($\sim5\cdot10^{29}$\,erg) is sufficient to sustain the chromospheric losses for more than two hours ($\sim5\cdot10^{27}$\,erg).

The spatial structure of, and the quantitative constraint on, the radiative losses will be of help to validate models that aim to reproduce reconnection in the lower solar atmosphere.   {Such characterization is crucial to understand reconnection processes and the triggering role of turbulent versus plasmoid reconnection. The advent of new-generation solar telescopes such as the Daniel K. Inouye Solar Telescope \citep[DKIST;][]{DKIST2015} and the European Solar Telescope \citep[EST;][]{EST2016} will be crucial in this regard.}

%%%%%%%%%%%%%%%%%%%%%%%%%%%%%%%%%%%%%%%%%%%%%%%%%%%%%%%%%%%%%%%%%%%%%%
\begin{acknowledgements}
%%%%%%%%%%%%%%%%%%%%%%%%%%%%%%%%%%%%%%%%%%%%%%%%%%%%%%%%%%%%%%%%%%%%%%

  {We would like to thank the anonymous referee for their comments and suggestions.}
  {CJDB thanks Sanja Danilovic and Flavio Calvo for their comments.}
%JdlCR
JdlCR is supported by grants from the Swedish Research Council (2015-03994), the Swedish National Space Agency (128/15), and the Swedish Civil Contingencies Agency (MSB). This project has received funding from the European Research Council (ERC) under the European Union's Horizon 2020 research and innovation program (SUNMAG, grant agreement 759548).
JL is supported through a project (CHROMATIC, 2016.0019) funded by the Knut and Alice Wallenberg foundation.
%SST
The Swedish 1-m Solar Telescope is operated on the island of La Palma by the Institute for Solar Physics of Stockholm University in the Spanish Observatorio del Roque de los Muchachos of the Instituto de Astrof\'isica de Canarias. The Institute for Solar Physics is supported by a grant for research infrastructures of national importance from the Swedish Research Council (registration number 2017-00625).
We acknowledge the community effort devoted to the development of the following open-source packages that were used in this work: numpy (\url{numpy.org}), matplotlib (\url{matplotlib.org}), scipy (\url{scipy.org}), astropy (\url{astropy.org}) and sunpy (\url{sunpy.org}).
Infrastructure for Computing (SNIC) at the PDC Centre for High Performance Computing (Beskow, PDC-HPC) at the Royal Institute of Technology in Stockholm as well as recourses at the National Supercomputer Centre (Tetralith, NSC) at Link\"oping University.
This research has made use of NASA's Astrophysics Data System Bibliographic Services.
\end{acknowledgements}

\bibliographystyle{aa}
% \bibliography{general}

\end{document}